\documentclass[journal=jacsat,manuscript=article]{achemso}
\usepackage{chemformula} 
\usepackage[T1]{fontenc} 

\author{Ning Ding}
\affiliation{School of Physics, Southeast University, Nanjing 21189, China}
\author{Kunihiro Yananose}
\affiliation{Center for Theoretical Physics, Department of Physics and Astronomy, Seoul National University, Seoul 08826, Republic of Korea}
\author{Carlo Rizza}
\affiliation{Department of Physical and Chemical Sciences, University of L’Aquila, Via Vetoio I-67100 Coppito, L’Aquila, Italy}
\author{Feng-Ren Fan}
\affiliation{Department of Physics, University of Hong Kong, Hong Kong, China}
\alsoaffiliation{HKU-UCAS Joint Institute of Theoretical and Computational Physics at Hong Kong, China}
\author{Shuai Dong}
\email{sdong@seu.edu.cn}
\affiliation{School of Physics, Southeast University, Nanjing 21189, China}
\author{Alessandro Stroppa}
\email{alessandro.stroppa@spin.cnr.it}
\affiliation{Consiglio Nazionale delle Ricerche, Institute for Superconducting and Innovative Materials and Devices (CNR-SPIN), c/o Department of Physical and Chemical Sciences, University of L'Aquila, Via Vetoio I-67100 Coppito, L'Aquila, Italy}
\alsoaffiliation{Department of Physical and Chemical Sciences, University of L’Aquila, Via Vetoio I-67100 Coppito, L’Aquila, Italy}

\title{Magneto-optical Kerr effect in ferroelectric antiferromagnetic two-dimensional heterostructures}
\abbreviations{IR,NMR,UV}
\begin{document}

\begin{tocentry}
    \centering
    \includegraphics[width=0.92\textwidth]{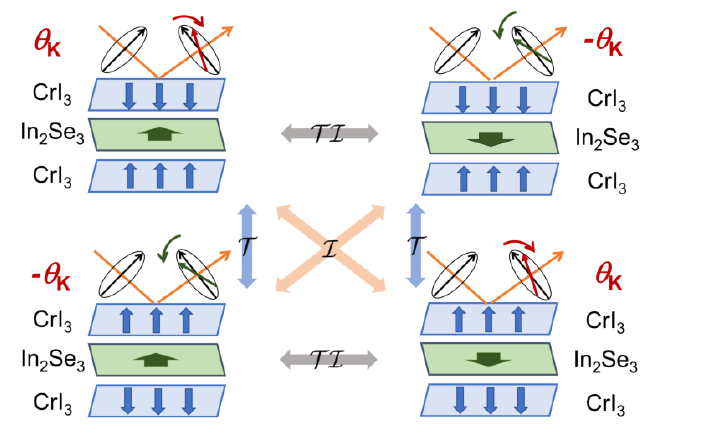}
\end{tocentry}

\maketitle
\begin{abstract}
 We study the magneto-optical Kerr effect (MOKE) of the two-dimensional heterostructure CrI$_3$/In$_2$Se$_3$/CrI$_3$  by using density functional theory calculations and symmetry analysis. The spontaneous polarization in the In$_2$Se$_3$ ferroelectric layer and the antiferromagnetic ordering in CrI$_3$ layers break the mirror symmetry and the time-reversal symmetry, thus activating MOKE. We show that the Kerr angle can be switched by either the polarization or the antiferromagnetic order parameter. Our results suggest that  ferroelectric and antiferromagnetic 2D heterostructures could be exploited for ultra-compact information storage devices, where the information is encoded by the two ferroelectric or the two time-reversed antiferromagnetic states, and the read-out performed optically by MOKE.
\end{abstract}

\section{Introduction}
Magneto-optical Kerr effect (MOKE) characterizes the change of the linear polarization of light reflected from a magnetized surface. It is usually active in ferromagnetic materials, where an external magnetic field or spontaneous ferromagnetic ordering  breaks the $\mathcal{T}$ (time-reversal) symmetry, leaving a net magnetization in the material.  The Kerr rotation is proportional to the sample magnetization so that MOKE is used to probe ferromagnetism especially for low-dimensional materials due to its harmlessness to samples and high sensitivity compared to traditional magnetic characterizations such as elastic neutron  scattering.\cite{gong2017-Nature,huang2017-Nature,Ribeiro2022-npj,maccaferri2015-acs-photonics}.
MOKE can be exploited also for magneto-optic information storage.\cite{tsunashima2001-JPD}

In general, a magnetic material can become magneto-optically active when both the $\mathcal{T}$ and $\mathcal{TI}$ symmetries are broken in the point group\cite{sivadas2016-PRL,yang2020-AEM}, where $\mathcal{I}$ is the spatial inversion. This suggests the way to turn on magneto-optical effects by breaking $\mathcal{I}$ symmetry since this automatically breaks $\mathcal{TI}$ symmetry, provided that $\mathcal{T}$ is already broken, like in magnetic systems. Since $\mathcal{I}$ is equivalent to $\mathcal{M}_z\mathcal{C}_2$, where $\mathcal{M}_z$ is the mirror reflection symmetry (mirror perpendicular to $z$-direction) and $\mathcal{C}_2$ is the in-plane inversion symmetry, the activation of MOKE can be achieved by breaking either $\mathcal{M}_z$ or $\mathcal{C}_2$, as confirmed by some theoretical and experimental reports\cite{sivadas2016-PRL,feng2015-PRB,feng2016-2D-M}. In general, MOKE is not active in antiferromagnetic systems, where the net magnetization is zero and the $\mathcal{TI}$ symmetry is usually preserved. However, several first-principles and experimental studies have predicted or verified a non-negligible MOKE in some antiferromagnets where the $\mathcal{TI}$ symmetry is broken by either the spatial asymmetry or chirality of crystal/spins even though the net magnetization is zero\cite{feng2015-PRB,yang2020-AEM,yang2022-PRB,feng2020-NC,zhou2021-PRB,higo2018-NPh,Fan2017-JACS,wu2020-APL,uchimura2022-APL}. In addition to its fundamental interest, the presence of MOKE in an antiferromagnet introduces additional perspectives, since, on the one hand, it challenges the traditional microscopic origin of the Kerr effect, on the other hand, it can be exploited in antiferromagnetic spintronics due to the insensitivity to external magnetic fields and ultrafast THz spin dynamics~\cite{jungwirth2016-Nat-Nano,Rustagi-PRB-2020,Yan-AM-2020,ma2022improving}.

Recently, two-dimensional (2D)  materials have been heavily studied since they show interesting physical properties different from bulk systems and can be exploited for miniaturized nanoelectronic components \cite{burch2018-Nature,gong2019-Science,zhang2022-NRM}. In particular, monolayer CrI$_3$, one of the first experimentally synthesized 2D magnets, has opened a new era in the study of 2D ferromagnetic materials~\cite{huang2017-Nature}. Soon after its discovery, many 2D magnetic materials have been theoretically predicted and experimentally investigated~\cite{gong2017-Nature,deng2018-Nature,ding2020-PRB,an2020-APL-Materials,you2022-AEM,chu2021-AM}. MOKE is a surface-sensitive probe for magnetism, thus  naturally attracting intensive interest from experimental and theoretical points of view for the study of magnetic 2D systems~\cite{gong2017-Nature,huang2017-Nature,wu2019-NC,jiang2022-PRB,gudelli2019-NJP,fang2018-PRB,molina2020-JMCC}. Recent studies have shown that MOKE can also be found in 2D antiferromagnetic materials~\cite{sivadas2016-PRL,yananose2022PRB,yang2020-AEM,wang2022-PRB}. 
	
Parallel to 2D magnets, also 2D ferroelectric materials, such as In$_2$Se$_3$~\cite{Ding2017-NC} and CuInP$_2$S$_6$~\cite{Liu2016-NC}, have received great attention. Inversion symmetry ($\mathcal{I}$) is  broken in ferroelectrics by the presence of spontaneous polarization. Therefore, the electric polarization or an external electric field in a layered collinear antiferromagnet may activate MOKE, since the  $\mathcal{TI}$ symmetry is also broken, providing the possibility to manipulate MOKE in antiferromagnetic ferroelectrics. For example, Sivadas et al. have recently shown that for layered collinear antiferromagnets, magneto-optic effects can be generated and manipulated through a gate voltage~\cite{sivadas2016-PRL}. This provides a promising route for electric field manipulation of the magneto-optic effects without modifying the underlying magnetic structure in antiferromagnetic heterostructure (HS). 2D HSs hold enormous potential for many applications since they can be assembled layer-by-layer to realize or to engineer desired physical properties \cite{Geim2013-Nature}.  

In this paper, we investigate a 2D HS composed by an In$_2$Se$_3$ ferroelectric monolayer sandwiched between two CrI$_3$ monolayers with antiferromagnetic coupling, thus forming a ferroelectric antiferromagnetic trilayer, such as CrI$_3$/In$_2$Se$_3$/CrI$_3$ HS. We use density functional theory (DFT) and symmetry analysis for the study of the structural, dynamical, ferroelectric and magneto-optical properties. Interestingly, the Kerr angle is reversed by the inversion of spontaneous polarization in In$_2$Se$_3$ and of the spins in CrI$_3$ monolayers, whereas it vanishes when the spontaneous polarization becomes zero. Furthermore, we investigate how the Kerr switching properties are related to the $k$-space spin textures of the system. Our results suggest the possibility of optically reading the information stored in  different polar antiferromagnetic states of the HS using MOKE. Therefore, 2D ferroelectric antiferromagnetic HS may be used for the development of novel and efficient magneto-optical devices in the field of antiferromagnetic spintronics\cite{wadley2016Science,winter2022PRB}.	
	
\section{Methods}
The first-principles calculations based on DFT have been performed using Vienna {\it ab-initio} Simulation Package (VASP) code~\cite{Kresse1996-PRB}, with the projector augmented wave method implementation in the generalized gradient approximation (GGA) parameterized by Perdew-Burke-Ernzerhof (PBE) for the exchange-correlation functional~\cite{kresse1999-PRB,Perdew1996-PRL}. The Hubbard $U_{\rm eff}$=$3$ eV with the Dudarev parameterization has been applied for properly describing the localization of Cr $3d$ orbitals~\cite{Dudarev1998-PRB}. A vacuum layer of $20$ \AA\ has been considered to avoid the interaction between the trilayer and their periodic images. The vdW correction for potential energy with DFT-D2 method has been used to describe the interlayer interaction among the three monolayers.\cite{Grimme2006-JCC} The $\Gamma$-centered {\it k}-point grid of $9 \times 9 \times 1$ has been employed to sample the Brillouin zone for the  static calculation and {\it k}-point grid is enlarged to $11 \times 11 \times 1$ for the dielectric function calculations. We have set the plane-wave cutoff energy to $450$ eV. The convergent criterion has been set to $10^{-5}$ eV for the energy, and to $0.01$ eV/\AA\ for the Hellman-Feynman forces during the structural optimization.
	
In order to evaluate the  energy stability, the binding energy ($E_b$) of the HS was calculated, i.e.
\begin{equation}
E_b=E({\rm HS})-E({\rm In_2Se_3})-2E({\rm CrI_3}),
\end{equation}
where $E({\rm HS})$, $E({\rm In_2Se_3})$, and $E({\rm CrI_3})$ are the energy of the HS, the  In$_2$Se$_3$ polar  monolayer, and CrI$_3$ ferromagnetic monolayer, respectively. The dynamic stability of the trilayer was evaluated based on the phonon spectra, using the finite differences methods\cite{phonopy,Page2002-PRB,Wu2005-PRB}. The polarization was calculated by the  Berry phase method~\cite{Resta1994-RMP,King1993-PRB}.

We have calculated the frequency-dependent relative dielectric tensor ($\underline{\underline{\epsilon}}$)red,~\cite{Gajdo2006-PRB} which was rescaled as shown in the method section in Supporting Information (SI). The relationship between $\underline{\underline{\epsilon}}$ and the effective surface optical conductivity tensor $\underline{\underline{\sigma}}$, describing the optical electromagnetic response of the 2D HS\cite{othman2013-OE}, is given by
\begin{equation}
\label{eq:epsilon_sigma}
\sigma_{jl}=-i d \epsilon_0 \omega [(\epsilon_{jl}-1)\delta_{jl}+\epsilon_{jl}(1-\delta_{jl})]
\end{equation}
where $j,l=x,y$, $\delta_{jl}$ is the Kronecker delta and $d$ is the trilayer longitudinal thickness, $\epsilon_0$ and $\omega$  are the vacuum permittivity and the radiation angular frequency, respectively.  

\section{The structures and symmetries}
Before studying the HS, we considered the individual properties of the isolated monolayers forming the HS. 
The optimized lattice constants of CrI$_3$ and In$_2$Se$_3$ monolayers are $7.064$ \AA\ and $4.101$ \AA, respectively. In addition, our calculations show that the CrI$_3$ and In$_2$Se$_3$ monolayers are semiconductors with a direct energy gap of $1.18$ eV and an indirect energy band gap of $0.66$ eV, respectively. The ferromagnetic moment of CrI$_3$ is $3$ $\mu_{\rm B}$/Cr and the polarization  of In$_2$Se$_3$ is about $2.98$ $\mu$C/cm$^2$ ($0.184$ e\AA\ per formula unit) by considering  the thickness of In$_2$Se$_3$ monolayer of $6.798$ \AA\ along the $z$-axis. The results are in good agreement with previous works \cite{mcguire2015-CM,Ding2017-NC,li2020-JAP,wu2019-NC}, thus supporting the accuracy and consistency of our calculations.

In order to model the HS, a $\sqrt{3}\times\sqrt{3}$ supercell of In$_2$Se$_3$ is used to match the unit cell of CrI$_3$, resulting in a very tiny ($0.55\%$) lattice mismatch. The stacking configuration of CrI$_3$/In$_2$Se$_3$/CrI$_3$ trilayer with lowest energy is shown in Fig.~\ref{fig1}(a), where the positions of chromium ions and selenium ions at top/bottom layer are overlapping along a common vertical axis with AB-stacking in CrI$_3$ bilayer\cite{sivadas2018-NL}, as depicted by the dashed lines in Fig.~\ref{fig1}(a). 

The space groups of the CrI$_3$ monolayer and AB-stacking CrI$_3$ bilayer are $P\bar{3}1m$ (No. 162) and $P\bar{3}$ (No. 147), respectively, where AB-stacking indicates the first layer laterally shifted by [$\frac{2}{3}$,$\frac{1}{3}$] in fractional coordinates to the second layer\cite{sivadas2018-NL}. For the In$_2$Se$_3$ monolayer, the space groups for ferroelectric state and paraelectric state are $P3m1$ (No. 156) and $P\bar{3}m1$ (No. 164), respectively\cite{Ding2017-NC}. However, the trilayer HSs with paraelectric and ferroelectric In$_2$Se$_3$ have a non-polar space group $P\bar{3}$ (No. 147) and a polar space group $P3$ (No. 143), respectively. 

The easy magnetization axis of CrI$_3$ monolayer is perpendicular to the layer plane (i.e. along the $c$-axis)\cite{huang2017-Nature}. Therefore, in the present work, the polar-MOKE geometry is considered, where the magnetization axis is along $z$-direction of the trilayer, i.e. perpendicular to the reflection surface and parallel to the plane of incidence. With this direction of the spin-moments, the magnetic space groups of paraelectric trilayer are $P\bar{3}$ (No. 147.13 in BNS setting) for ferromagnetic CrI$_3$ bilayer and $P\bar{3}'$ (No. 147.15 in BNS setting) for layered antiferromagnetic CrI$_3$, which correspond to the magnetic point groups (MPG) of $\bar3$ and $\bar{3}'$, respectively. However, the magnetic space group of the ferroelectric trilayer is $P3$ (No. 143.1 in BNS setting) regardless of whether the magnetic coupling between the two CrI$_3$ layers is ferromagnetic or antiferromagnetic.     

The traditional approach to evaluate the Kerr rotations from first-principles calculations relies on a well-known formula, which is valid in three-fold rotational symmetry and suitable for the case where one is not interested in surface specific phenomena\cite{ebert1996-RPP,sangalli2012-PRB,rosa2015-PRB}. However, the finite thickness in 2D materials limits the application of this approach. Therefore, we provide an extension in order to take into account the finite thickness of the HS. We consider a monochromatic plane wave normally impinging on a 2D HS on a dielectric substrate and whose electric field is given by ${\bf E}=Re \left( {\bf E}_{\omega} e^{-i \omega t}  \right)$, where ${\bf E}_{\omega}={E}_{\omega,x} \hat{\bf e}_x+{E}_{\omega,y} \hat{\bf e}_y$. Considering that the HS lies on the plane $z=0$ and its longitudinal size is much smaller than the radiation wavelength, the HS electromagnetic response can be effectively described by the optically induced surface current density ${\bf K}=\underline{\underline{\sigma}} {\bf E}_{\omega}$, where $\underline{\underline{\sigma}}$
is the surface conductivity tensor, i.e. a second-rank tensor.~\cite{yoshino2013-JPSA} Clearly, the form of $\underline{\underline{\sigma}}$ strongly depends on symmetry. 
For a paraelectric trilayer, the MPG is $\bar{3}'$ and thus the surface optical conductivity tensor $\underline{\underline{\sigma}}$ is a diagonal tensor where $\sigma_{xx}=\sigma_{yy}$. However, due to the MPG of the ferroelectric trilayer ($3$), $\underline{\underline{\sigma}}$ can be written as:\cite{stokes2005findsym,gallego2019-ACSA} 
\begin{equation}
\label{sigma}
\underline{\underline{\sigma}}=
\begin{pmatrix}
\sigma_{xx} & \sigma_{xy}  \\
-\sigma_{xy} & \sigma_{xx}  \\
\end{pmatrix}. 
\end{equation}
Next, one can obtain the rigorous reflection coefficients by solving Maxwell's equations and requiring that the electric and magnetic ($\bf H$) fields fulfill the boundary conditions in the presence of the surface current $\bf K$, namely
\begin{eqnarray}
&& \left(\bf{E}_{\omega}^{>}-\bf{E}_{\omega}^{<} \right)=0  \nonumber \\
&& \hat{\bf e}_z \times \left(\bf{H}_{\omega}^{>}-\bf{H}_{\omega}^{<} \right)=\bf{K}, 
\end{eqnarray}
where $>$ and $<$ indicate the upper ($z>0$)and lower $(z<0)$ sides of the 2D HS, respectively. The reflection matrix connecting the component of the 
incident (${\bf E}_{\omega}^{(i)}$) and reflected (${\bf E}_{\omega}^{(r)}$) electric fields is: 
\begin{eqnarray}
\begin{pmatrix}
& E_{\omega,x}^{(r)} \\ & E_{\omega,y}^{(r)}   
\end{pmatrix}=
\begin{pmatrix}
r_{xx} & r_{xy}  \\
r_{yx} & r_{yy}  \\
\end{pmatrix}
\begin{pmatrix}
&E_{\omega,x}^{(i)} \\ & E_{\omega,y}^{(i)}  
\end{pmatrix}, 
\end{eqnarray}
where 
\begin{eqnarray}
  		&& r_{yy}=r_{xx}=\frac{1-(\xi_{xx}+{n_{s}})^2-\xi_{xy}^2}{(n_{s}+1+\xi_{xx})^2+\xi_{xy}^2} \nonumber \\
  		&& r_{xy}=-r_{yx}=\frac{2\xi_{xy}}{(n_{s}+1+\xi_{xx})^2+\xi_{xy}^2},
\end{eqnarray}
dimensionless optical conductivity $\xi_{ij}$ = $\sigma_{ij} \cdot Z_0$ ($Z_0$ is the vacuum impedance), and $n_s$ is the substrate refractive index.
The complex Kerr angle $\phi$ can be defined as $\phi_{\text{K}}=r_{xy}/r_{xx}=\theta _{\text{K}} +i\eta _{\text{K}}$ in terms of the rotation angle $\theta _{\text{K}}$ and ellipticity $\eta _{\text{K}}$~\cite{ricci2008-PRB}, so that 
\begin{equation}
\label{phi}
\theta _{\text{K}} +i\eta _{\text{K}}=\frac{2\xi_{xy}}{1-(\xi_{xx}+{n_{s}})^2-\xi_{xy}^2}
\end{equation}
From Eq.(\ref{phi}), it is evident that MOKE is activated for $\xi_{xy} \neq 0$.~\cite{yang2020-AEM} Before calculating the Kerr angle of HS, we confirmed that of CrI$_3$ monolayer was comparable with experimental one as depicted in the Figure S1 of SI, which can ensure the accuracy of the following calculation.

Here, we focus our attention on the antiferromagnetic configuration and how the MOKE activity is induced by the electric polarization of the In$_2$Se$_3$ monolayer. In order to describe the system response, we introduce two order parameters, i.e., $\mathbf{L} = \mathbf{M}_1-\mathbf{M}_2$ (the antiferromagnetic order parameter) and $\mathbf{P}$ (ferroelectric order parameter), where the labels $1$ and $2$ correspond to the bottom and top layers of CrI$_3$ bilayer, respectively. Here, $\mathbf{M}$ and $\mathbf{P}$ are the spin magnetic moment and electric polarization collinear to the $c-$axis, which is oriented from bottom to top layer. One can focus on four different states of ferroelectric trilayer antiferromagnets, such as ($+$L,$+$P), ($+$L,$-$P), ($-$L,$+$P) and ($-$L,$-$P), respectively, where +/$-$L,+/$-$P are the projections of $\mathbf{L}$ and $\mathbf{P}$ along the $c$-axis. These four states can be related by the spatial inversion symmetry ($\mathcal{I}$) and time-reversal symmetry ($\mathcal{T}$), i.e., $\mathcal{I}$$\mathbf{M}$=$\mathbf{M}$ and $\mathcal{I}$$\mathbf{P}$=-$\mathbf{P}$, while $\mathcal{T}$$\mathbf{M}$=-$\mathbf{M}$ and $\mathcal{T}$$\mathbf{P}$=$\mathbf{P}$. We summarize the action of the symmetry operations on our system in Fig.~\ref{fig1} (b), and the symmetry relations of the four states are shown in Fig.~\ref{fig1} (c). For example, the ($+$L,$+$P) can turn to ($+$L,$-$P) by $\mathcal{TI}$ symmetry operation.
	
\section{Results and discussion}

\begin{figure}[t!]
\includegraphics[width=0.7\textwidth]{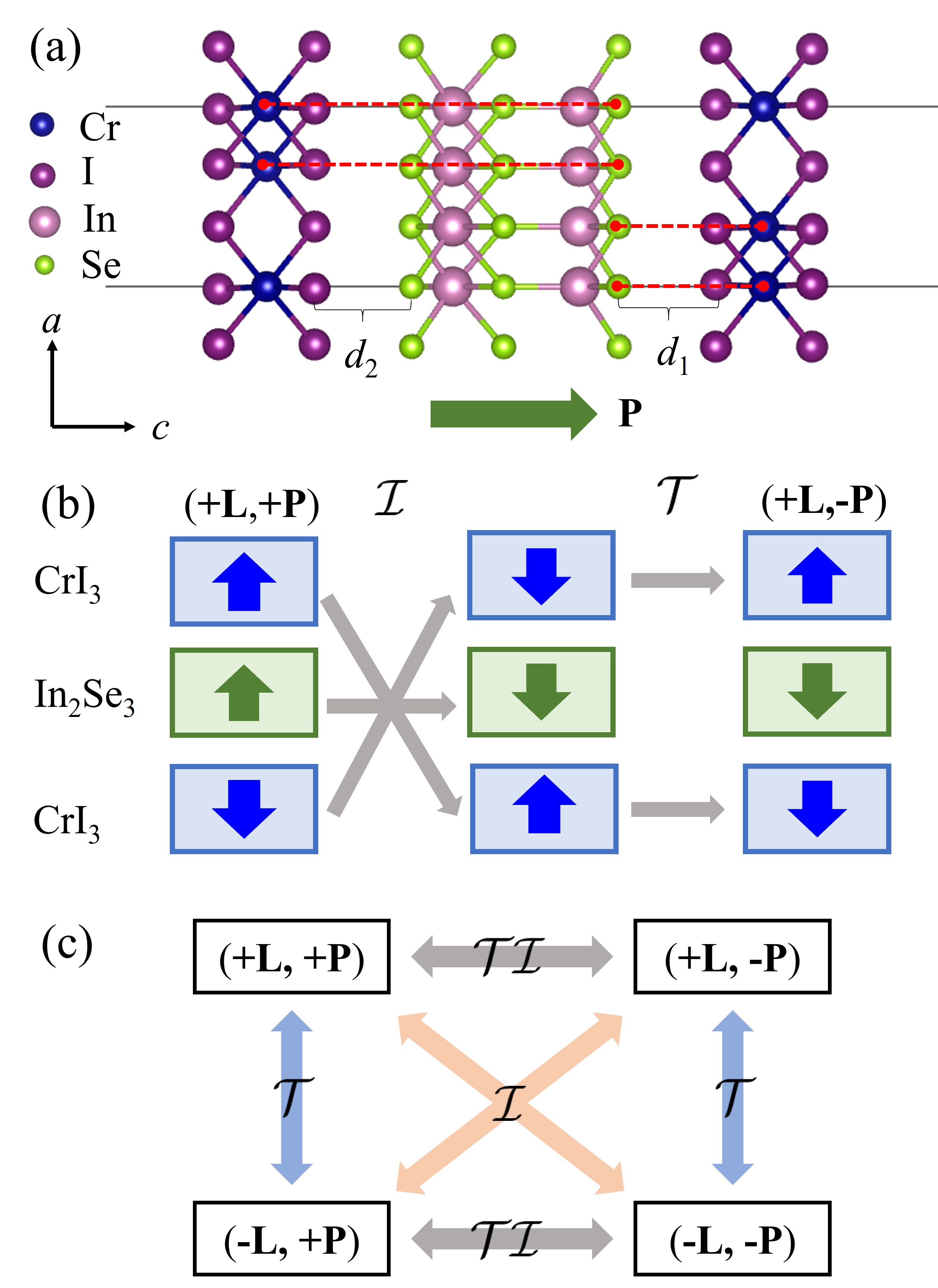}
\caption{(a) Side view of the ferroelectric-antiferromagnetic trilayer structure, where the green arrow indicates the polarization direction of In$_2$Se$_3$ monolayer. (b) Schematic rule of $\mathcal{I}$ and $\mathcal{T}$ switching in the trilayer, where the blue arrows in CrI$_3$ denote the spins and the green arrows in In$_2$Se$_3$ indicate the ferroelectric polarization. The consequences after each of the $\mathcal{I}$ and $\mathcal{T}$ operations are shown as grey arrows. (c) The symmetry relations of four states. $\textbf{L}$: antiferromagnetic order parameter; $\textbf{P}$: ferroelectric order parameter (i.e. the polarization). The symmetry operations are represented by double-sided arrows.}
\label{fig1}
\end{figure}

\begin{figure}[t!]
\includegraphics[width=0.5\textwidth]{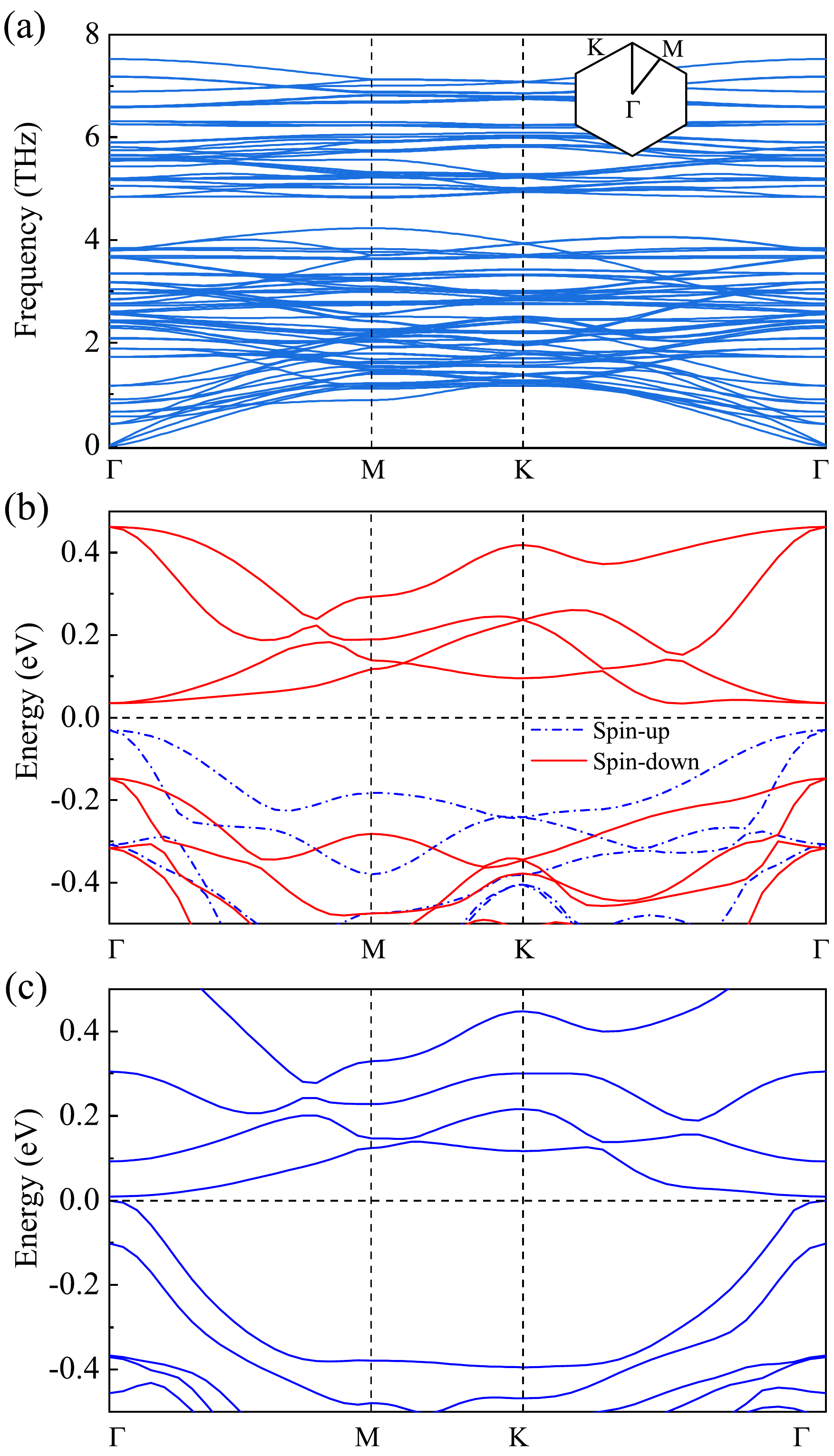}
\caption{(a) The phonon dispersion curves of ferroelectric trilayer antiferromagnets. The high symmetry points are defined in the inset. (b-c) The band structures of the ferroelectric trilayer antiferromagnets (b) with the spin polarization but without SOC and (c) with SOC.}
\label{fig2}
\end{figure}

The HS shows three stacking modes with three-fold rotational symmetry. Moreover, the system can have different magnetic orders (see SI for further details). Here, we focus on the most energetically favored state reported in Fig.~\ref{fig1}(a). In this case, the  lattice constant of the ferroelectric trilayer with antiferromagnetic coupling is {\it a} = {\it b} = $6.905$ \AA. The optimized interlayer spacing {\it d}$_1$ and {\it d}$_2$ are both $3.218$ \AA, while the longitudinal thickness ($d$) is $19.711$ \AA. The energy of the antiferromagnetic state in the trilayer system is about $0.11$ meV/atom  higher than that of the ferromagnetic state, i.e. almost degenerate with it. However, the relative energy is very sensitive to computational details, and, in addition, it may depend on the experimental details during the synthesis or preparation of the system. The magneto-optical response and its switching properties are straightforward in the case of ferromagnets,  so, for the purpose of the present study, we choose to investigate the antiferromagnetic configurations. We calculated  the thermodynamic stability by evaluating the binding energy ($E_b$) of the HS with antiferromagnetic coupling, which is about $-44.51$ meV/atom. This value is comparable to $E_b$ evaluated in previous work~\cite{Liu2022-Nanoscale}. For the dynamic stability, we have evaluated the phonon dispersions. As shown in Fig.~\ref{fig2}(a), there is no  imaginary frequency within the numerical uncertainty, indicating the dynamic stability of this system.

\begin{figure}[t!]
\includegraphics[width=0.7\textwidth]{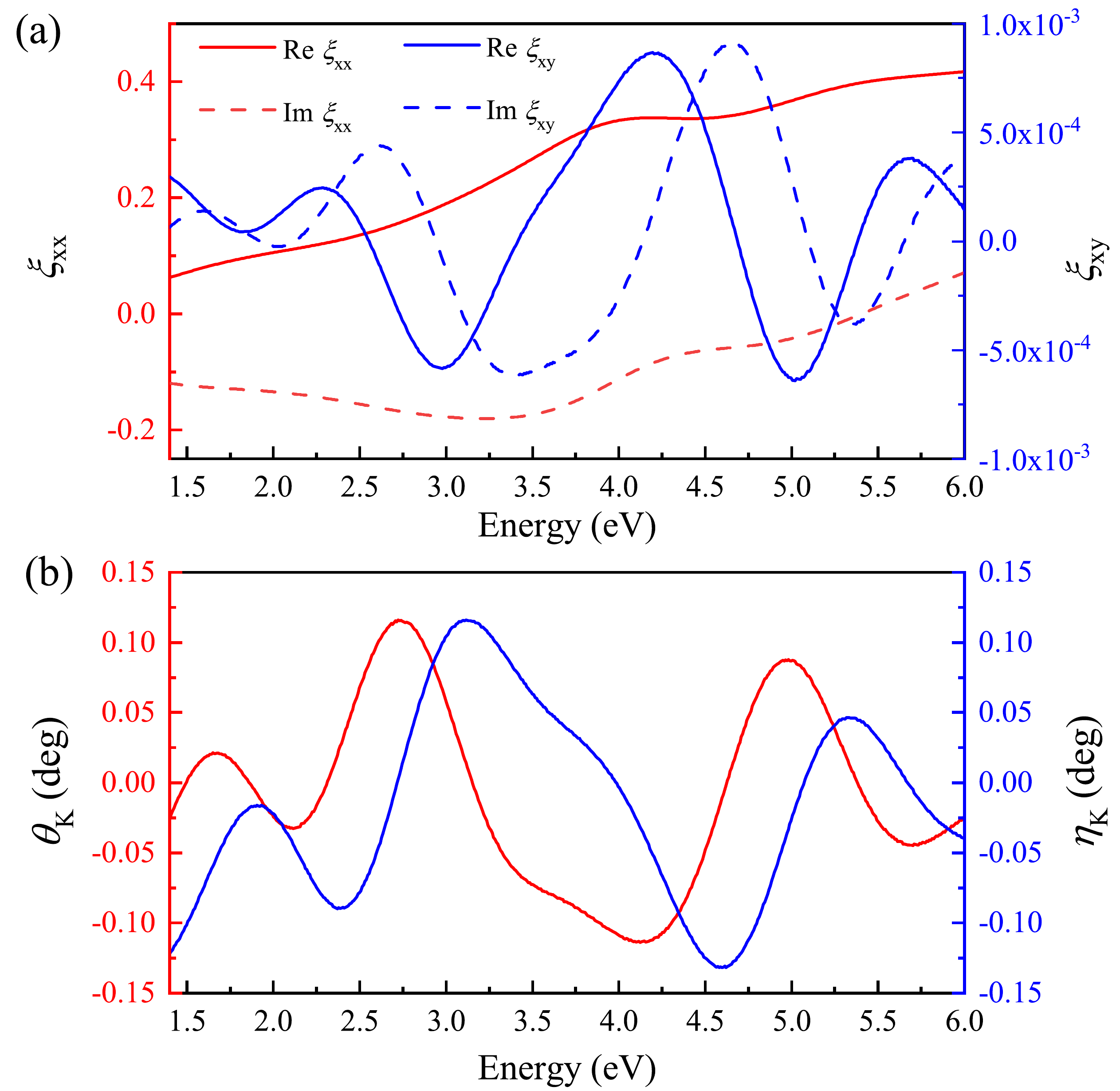}
\caption{(a) The frequency-dependent dimensionless surface optical conductivity of the CrI$_3$/In$_2$Se$_3$/CrI$_3$ HS in the polar state. (b) Kerr angle $\theta_{\text{K}}$  and ellipticity $\eta_{\text{K}}$ of Eq.(\ref{phi}) using the dimensionless surface optical conductivity of panel (a) and setting $n_s=1$.}
\label{fig3}
\end{figure}

Although the CrI$_3$ bilayer is antiferromagnetic (a nearly zero net magnetic moment of $0.001$ $\mu$$_{\text{B}}$/Cr), the spin-up and spin-down bands are not degenerate due to the broken $\mathcal{TI}$ symmetry caused by the existence of polarization, as shown in Fig.~\ref{fig2}(b). The effect is reminiscent of bilayer MnPSe$_3$ gated by electric field\cite{sivadas2016-PRL}. In our work, the existence of polarization at +$P$ state induces the conduction band minimum (CBM) of spin-down (spin-up) channel of CrI$_3$ monolayer downward (upward) shift, and this induces the spin-up channel exhibits a much wider band gap than the spin-down channel, where the spin-up (-down) corresponds to the bottom (top) CrI$_3$ layer. Therefore, the band gap reduces, which is a common phenomenon in vdW HS system.\cite{chen2021-NC,qiu2019-SA,zhang2018-Nanoscale} The band-splitting in non-centrosymmetric antiferromagnet correlates with the activation of MOKE, which can play the role of exchange splitting. On the other hand, the band structure of the paraelectric trilayer, shown in Fig. S3, does not exhibit spin-polarized bands splitting, which would correlate with a MOKE inactivity in the paraelectric trilayer system. The band structure with the spin-orbit coupling (SOC) is further split as reported in Fig.~\ref{fig2}(c).  
	
The easy magnetization axis of the ferroelectric antiferromagnetic trilayer is along the {\it z}-axis with the magnetic anisotropy energy of $0.56$ meV/Cr. The spin-configuration of the trilayer is analyzed using the FINDSYM resulting in an  MPG $3$ as expected\cite{stokes2005findsym,stokes2005findsym}. Furthermore, according to MTENSOR tool\cite{gallego2019-ACSA}, the off-diagonal component of dimensionless surface optical conductivity $\xi_{xy}$ should be non-zero in this group as shown in Fig.~\ref{fig3}(a). Therefore, the Kerr angle of the system has been investigated in the $1.4-6.0$ eV photon energy range  which should be more relevant for experiments, as depicted in Fig.~\ref{fig3}(b).
	
\begin{figure}[t!]
\includegraphics[width=0.8\textwidth]{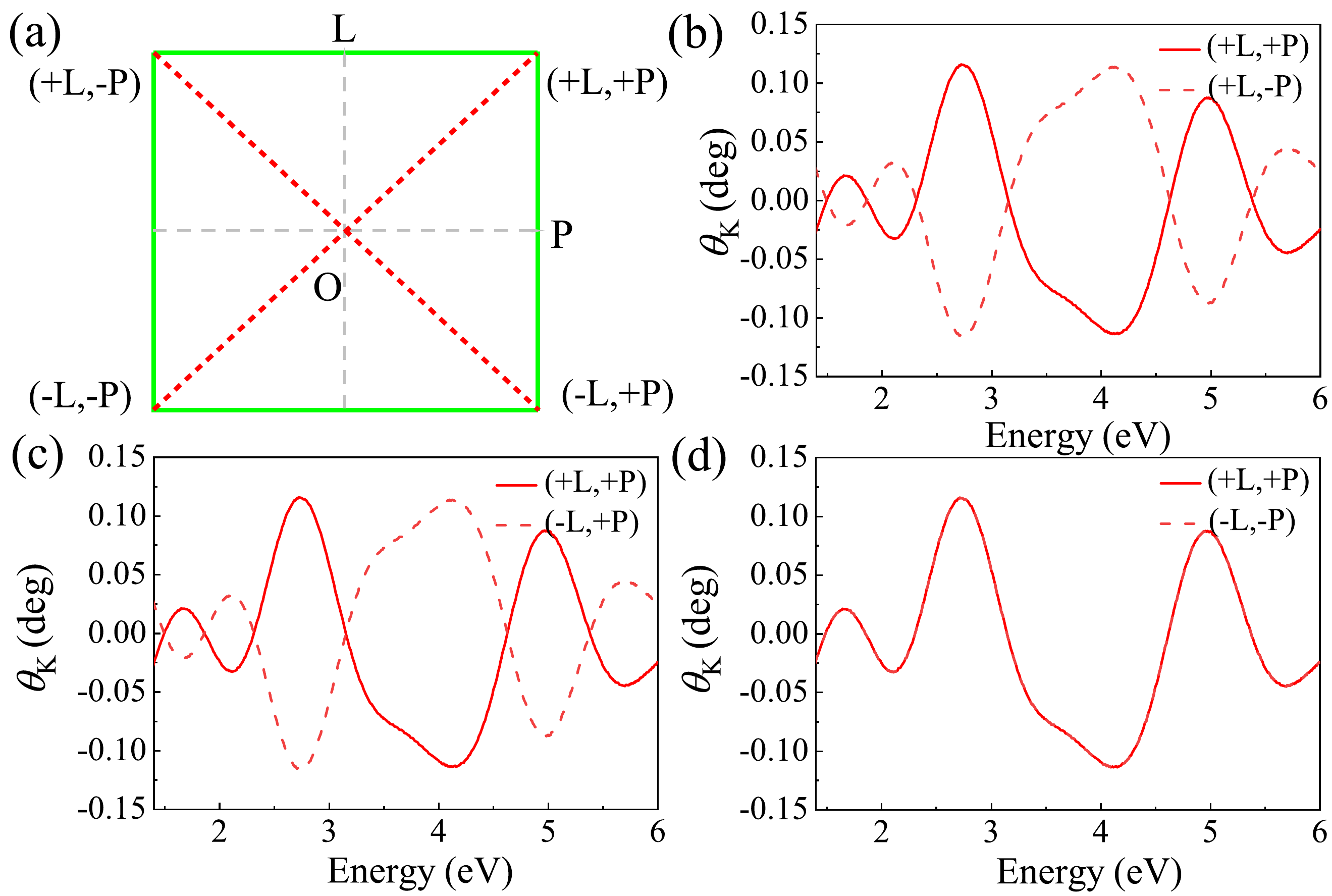}
\caption{(a) Representation of the dependence of Kerr angle of the order parameters $P$ and $L$ on a ($P$,$L$) plane.  The green solid (red dotted) lines represent the switchability (non-switchability) of the Kerr angle. (b)-(d) Calculated Kerr angle of four states, the solid lines represent (+L,+P), and the dashed lines represent ($-$L,+P), ($-$P,+L) and ($-$L,$-$P), respectively.}
\label{fig4}
\end{figure}
	
Using the transformation rules of the optical conductivity tensor under  $\mathcal{I}$ and $\mathcal{T}$ symmetry operations, we note that  under the $\mathcal{I}$ symmetry operation, $\underline{\underline{\sigma}}$ of MPG $3$ transforms as follows,
\begin{equation}
\mathcal{I}:\underline{\underline{\sigma}}\mapsto \mathcal{I}\underline{\underline{\sigma}} \mathcal{I}^{-1} =\underline{\underline{\sigma}},
\end{equation}
where $\underline{\underline{\sigma}}$ is invariant under inversion operation $\mathcal{I}$. This implies that the Kerr angle is also invariant under inversion in the ferroelectric antiferromagnetic HS.
	
On the other hand, under the $\mathcal{T}$ symmetry operation, $\underline{\underline{\sigma}}$ of MPG $3$ transforms as follows,
\begin{equation}
\mathcal{T}:\underline{\underline{\sigma}}\mapsto \mathcal{T}\underline{\underline{\sigma}} \mathcal{T}^{-1}=
\underline{\underline{\sigma}}^{t},
\end{equation}
where `$t$' means the transpose of the conductivity tensor according to Onsager's relation~\cite{Rathgen2005-PRB}. The $\sigma_{xy}$ as shown in Eq. (\ref{sigma}) changes the sign under $\mathcal{T}$ operation, thus implying that the Kerr angle is also inverted under $\mathcal{T}$ operation according to Eq. (\ref{phi}). In the same way, it can be shown that the $\mathcal{TI}$ operation can also reverse the Kerr angle in MPG $3$ system.
	
It is interesting to investigate the switching rule for the Kerr angle among the four states according to the symmetry relations as depicted in Fig.\ref{fig1}(c) and Eq.(\ref{phi}), which can be summarized as follows. The Kerr angle can be reversed: (a) 
by either the inversion of polarization $\mathbf{P}$ or (b) the inversion of antiferromagnetic parameter $\mathbf{L}$, in addition, 
(c) the Kerr angle is invariant inverting both $\mathbf{P}$ and $\mathbf{L}$. This is summarized in Fig.~\ref{fig4}(a), where we represent the four relevant states on a ($P$,$L$) plane, and the switchability (non-switchability) of the Kerr angle is associated  to  green solid (red dotted) lines. As shown in Fig.~\ref{fig4}(b-d), these predictions based on symmetry are further proved by the first-principles calculations.

\begin{table}
\caption{The simulated  displacements ($\delta$) deviating from ferroelectric state along $z$ direction and the corresponding polarization ($P$). The thickness of HS is $d$ = $19.711$ \AA.}
\centering
\begin{tabular*}{0.7\textwidth}{@{\extracolsep{\fill}}lccccc}
\hline \hline
$\delta$ (\AA) & $-0.08$ & $-0.04$ & $0$ & $0.04$ & $0.08$ \\
\hline
$P$ (e\AA) & $0.459$ & $0.536$ & $0.573$ & $0.610$ & $0.687$ \\
$P$ ($\mu$C/cm$^2$) & $0.904$ & $1.055$ & $1.128$ & $1.201$ & $1.352$ \\
\hline \hline
\end{tabular*}
\label{table1}
\end{table}

\begin{figure}[t!]
\includegraphics[width=0.7\textwidth]{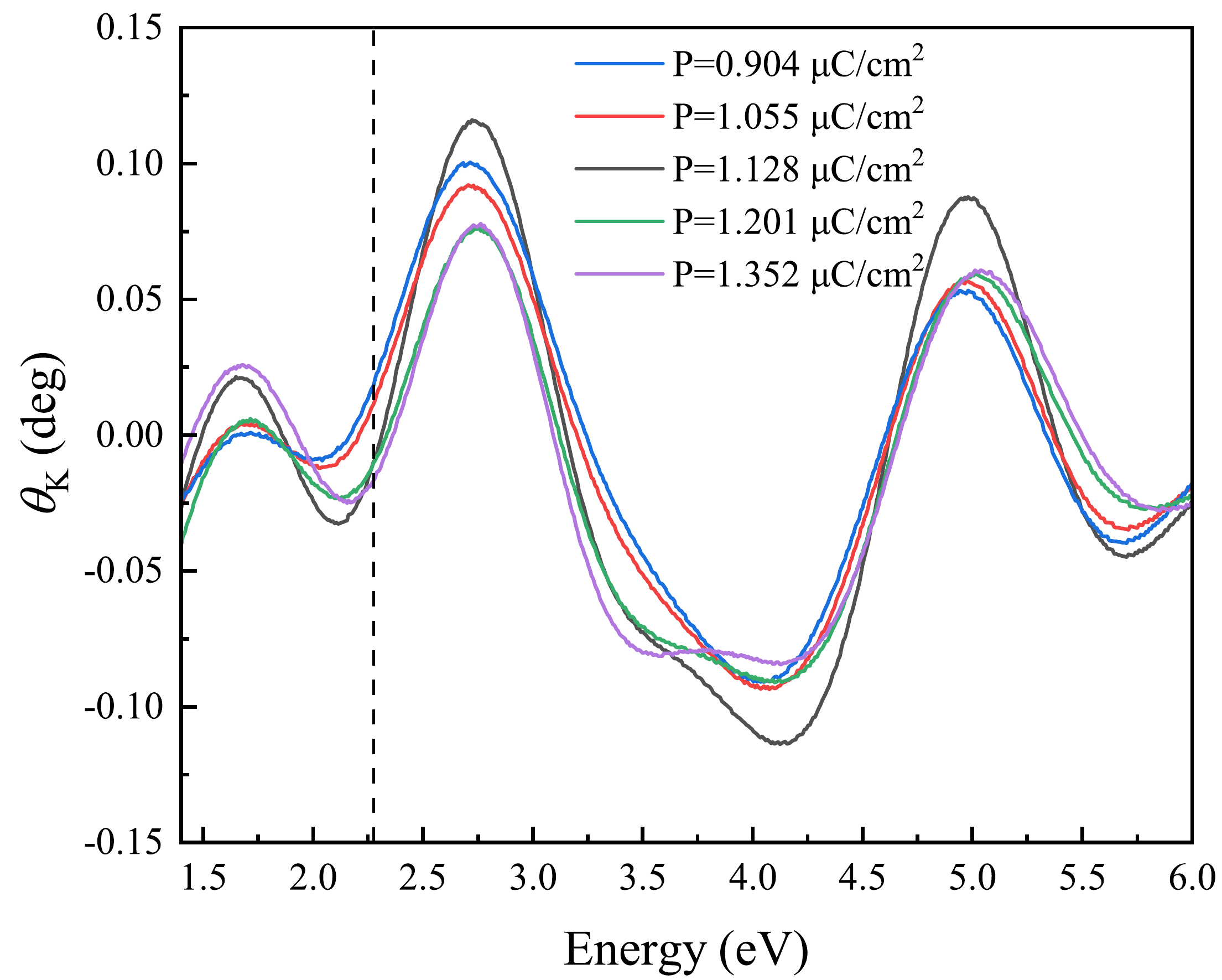}
\caption{Kerr angle $\theta_{\text{K}}$ as a function of the spontaneous polarization $P$ in the considered ferroelectric antiferromagnetic HS (i.e., CrI$_3$/In$_2$Se$_3$/CrI$_3$).}
\label{fig5}
\end{figure}
 
Qualitatively, the switching behavior of the Kerr angle can be expressed as follows: $\theta_{\text{K}}(P,L;E) = \alpha PL$, where $\alpha$ is a function of $E$. Indeed, the sign change of L or P, implies the switching of  $\theta_{\text{K}}$, 
but the simultaneous sign change of L and P will not affect $\theta_{\text{K}}$. Furthermore, the formula suggests that
different materials, having a different magnitude  of $P$ or $L$, 
can give different values of Kerr angle at a specific photon energy. Note that the coefficient $\alpha$ is energy- and material-dependent. In order to explore this possibility, we perform a series of computational experiments tuning the polarization by artificially  shifting the position of the intermediate selenium atoms in the $z$ direction. For the sake of convenience, this artificial displacement ($\delta$) is defined  with respect to the  atomic positions in the ferroelectric state, as shown in Table \ref{table1}, thus affecting the final polarization. The results of the Kerr angle with different polarization are shown in Fig.~\ref{fig5}. Interestingly, the Kerr angle exhibits a monotonous relation with $P$ at a specific energy value. For example, the Kerr angle at $2.3$ eV increases as the polarization increases.

The physical mechanism can be explained by the different spin-splitting with different polarization as shown in the Fig.S6 in the SI. For example, larger magnitude of polarization results larger downward shift of CBM of  spin-down channel of CrI$_3$ monolayer, and this enlarges the spin splitting, furthermore, it causes a larger Kerr angle and vice versa. These results suggest the possibility of optimizing the Kerr response for a specific energy range by choosing a suitable ferroelectric and antiferromagnetic composite system.

\begin{figure}[t!]
\centering
\includegraphics[width=0.4\linewidth]{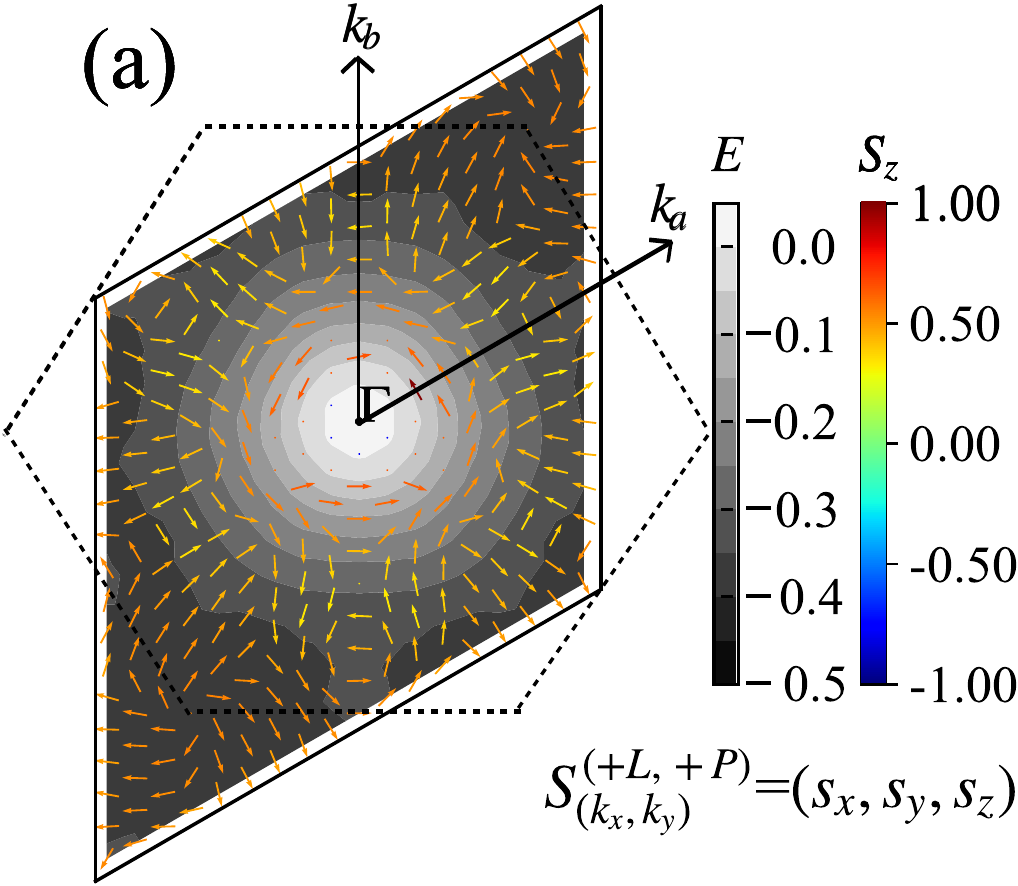}
\hspace{0.01\linewidth}
\includegraphics[width=0.4\linewidth]{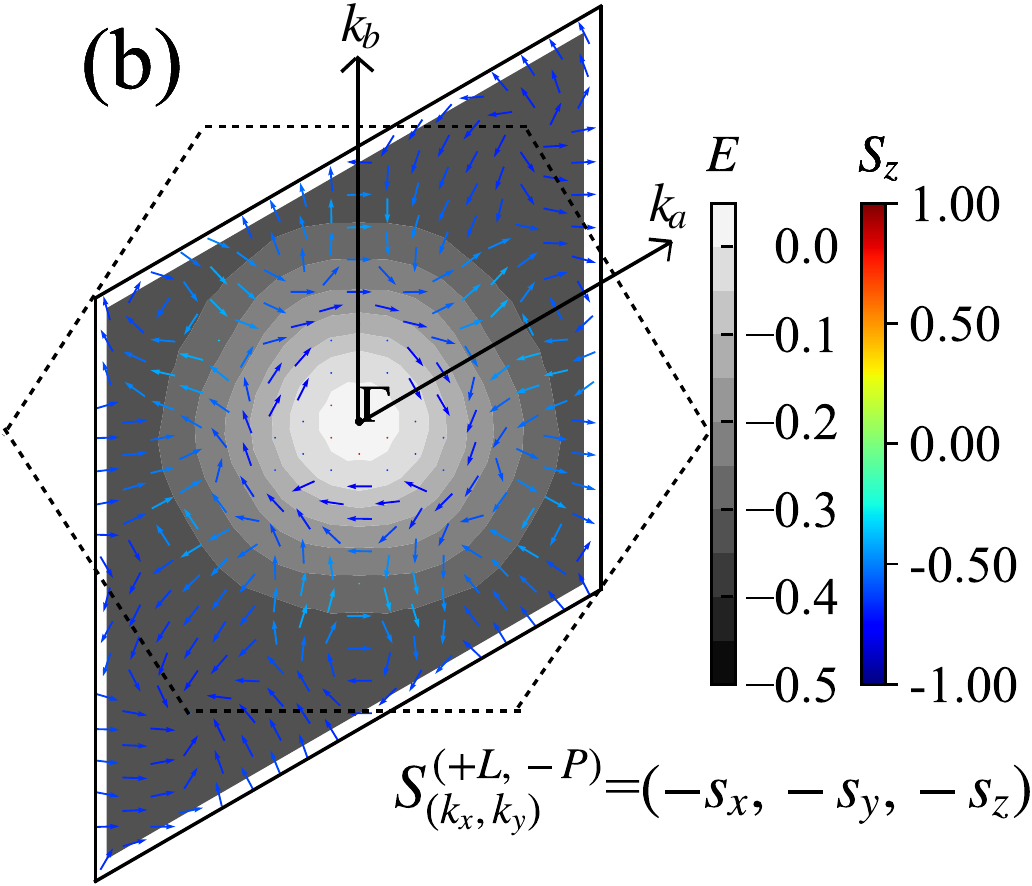}
\vfill
\includegraphics[width=0.4\linewidth]{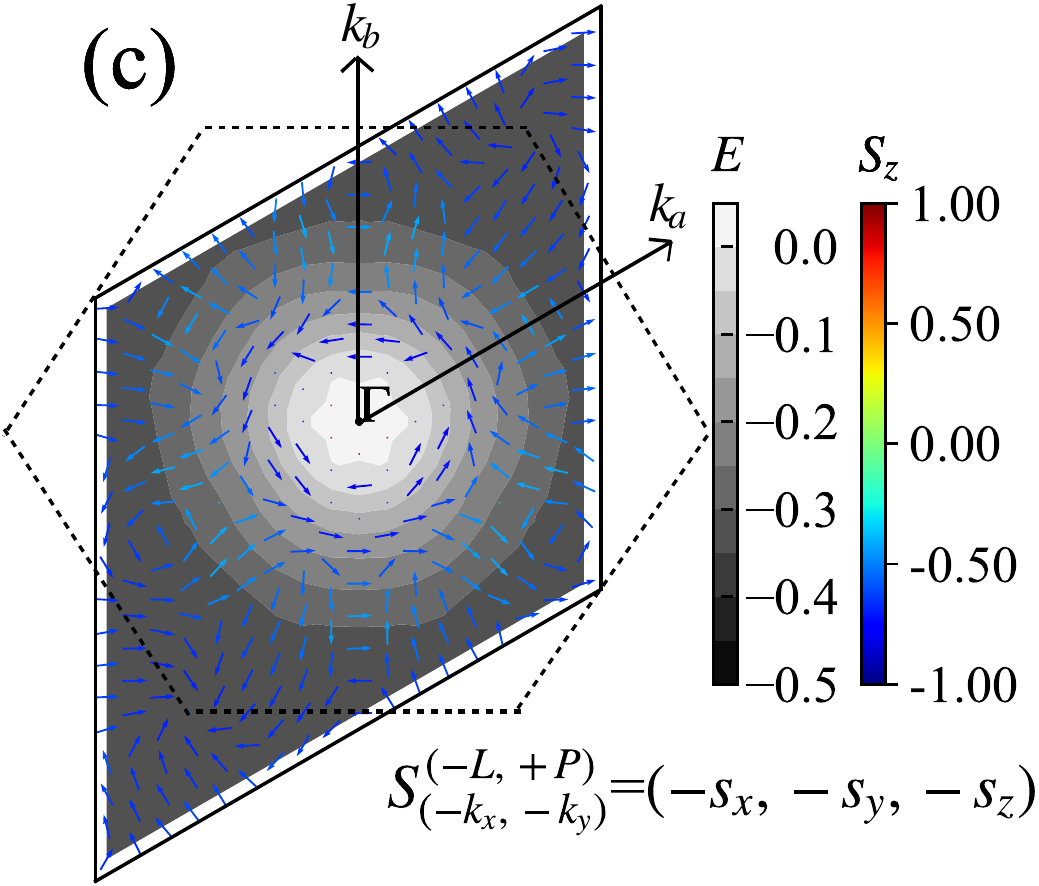}
\hspace{0.01\linewidth}
\includegraphics[width=0.4\linewidth]{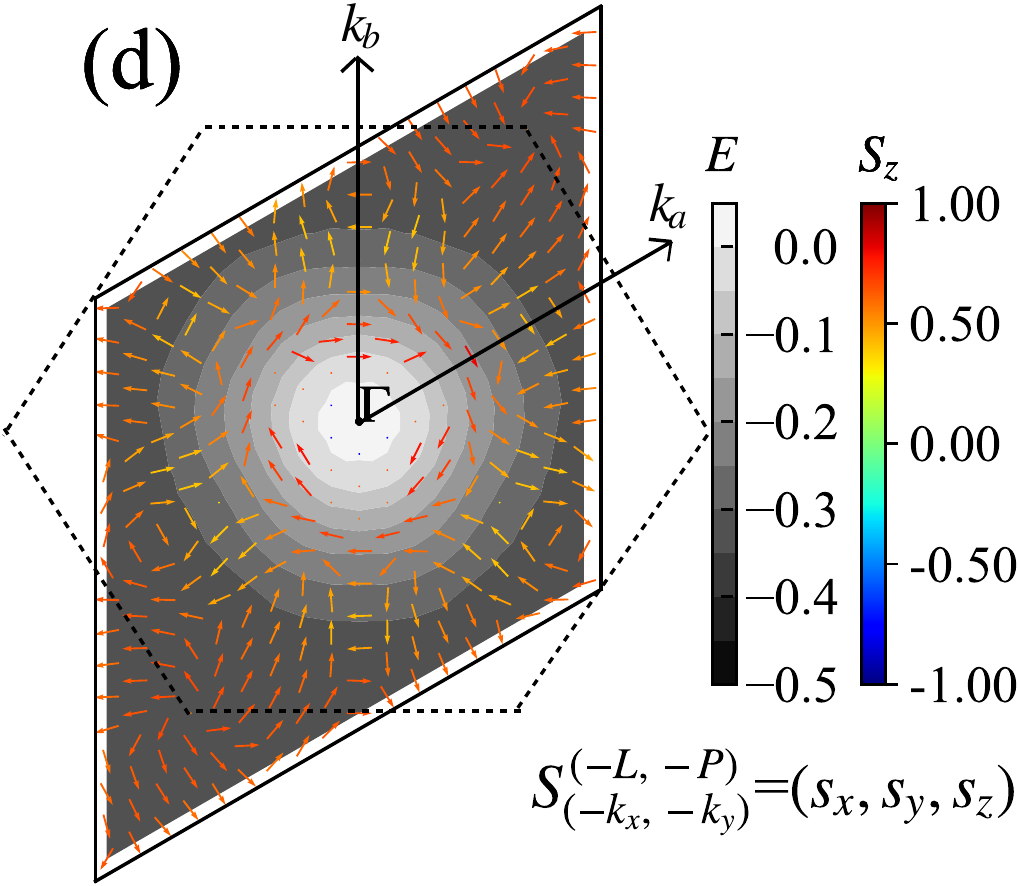}
\vfill
\includegraphics[width=0.85\linewidth]{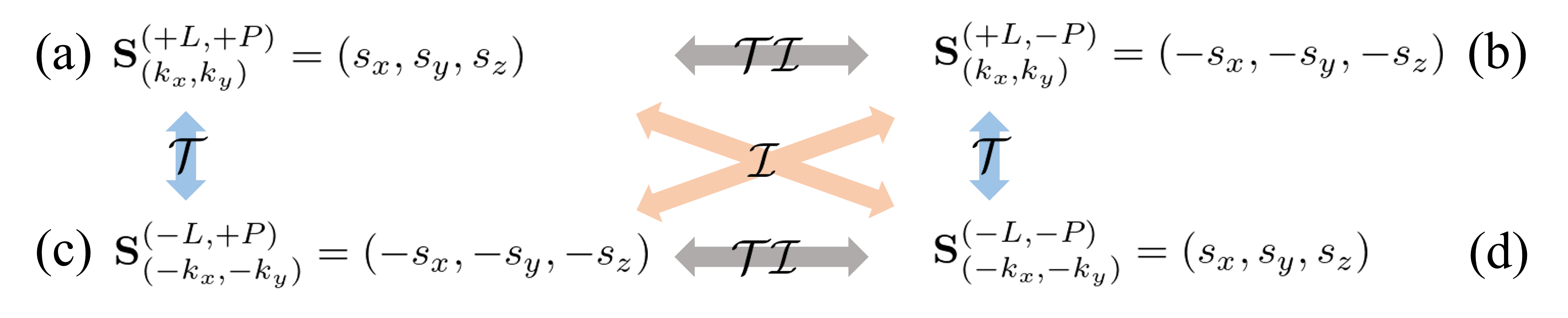}
\caption{The spin texture, where the in-plane spin components are represented by an arrow and the out-of-plane components are shown in color-map, and its relationship of four states for the highest valence band. (a) (+L,+P), (b) ($-$L,+P), (c) (+L,$-$P), (d) ($-$L,$-$P). The first Brillouin Zone (1$^\text{st}$ BZ) can be shown by the diamond, which is equivalent to a regular hexagon (the common choice of 1$^\text{st}$ BZ of hexagonal lattice as depicted by the black dotted lines) by shifting its fragments. The energy level with respect to the valence band maximum are illustrated by grey-scale map. The figure at the bottom describes intuitively the relationship between the spin-texture and symmetry operations.}
\label{fig6}
\end{figure}
 
Finally, we consider how the switching rules discussed for the MOKE response correlate with the spin texture for the highest valence bands in {\it k}-space for the four different states. We summarize the results in Fig.~\ref{fig6}. Clearly, the spin texture for the four states has the three-fold rotational symmetry, consistent with the point group  ${\it C}_{3}$, and, accordingly, the  spin texture should have a Rashba-type topology~\cite{Tao_2021-JPD}. However, the SOC splitting in this work exhibits a different behavior from the conventional Rashba band, there is no linear crossing due to the non-existence of Kramers degeneracy caused by the time-reversal breaking. However, SOC splitting ($\Delta$ $\approx$ $0.10$ eV) for the highest valence band at the $\Gamma$ point can be used to indirectly characterize the magnitude of Rashba parameter, which is compared to that at the Y point ($\Delta$ $\approx$ $0.09$ eV) of BiInO$_3$~\cite{tao2018-NC}. The transformation rules for spin and {\it k} by the $\mathcal{I}$, $\mathcal{T}$, and $\mathcal{TI}$ operations are as follows. For $\mathcal{T}$, (${\it s}_{x}$, ${\it s}_{y}$, ${\it s}_{z}$) $\longmapsto$ ($-{\it s}_{x}$, $-{\it s}_{y}$, $-{\it s}_{z}$) and (${\it k}_{x}$, ${\it k}_{y}$) $\longmapsto$ ($-{\it k}_{x}$, $-{\it k}_{y}$). For $\mathcal{I}$, (${\it s}_{x}$, ${\it s}_{y}$, ${\it s}_{z}$) $\longmapsto$ (${\it s}_{x}$, ${\it s}_{y}$, ${\it s}_{z}$) and (${\it k}_{x}$, ${\it k}_{y}$) $\longmapsto$ ($-{\it k}_{x}$, $-{\it k}_{y}$). Combining those two effects, for $\mathcal{TI}$, (${\it s}_{x}$, ${\it s}_{y}$, ${\it s}_{z}$) $\longmapsto$ ($-{\it s}_{x}$, $-{\it s}_{y}$, $-{\it s}_{z}$) and (${\it k}_{x}$, ${\it k}_{y}$) $\longmapsto$ (${\it k}_{x}$, ${\it k}_{y}$). Since the four states are connected by the $\mathcal{T}$, $\mathcal{I}$, or $\mathcal{TI}$ operations, as summarized in  Fig.~\ref{fig6}, it follows that  the spin-texture can be reversed either by P or L reversal, as it happens for the Kerr rotation.  

\section{Conclusions}
 We have investigated the 2D ferroelectric antiferromagnetic HS CrI$_3$/In$_2$Se$_3$/CrI$_3$ by focusing on its magneto-optical Kerr effect response. Interestingly, in the antiferromagnetic configuration, MOKE is activated and controlled by the spontaneous electric polarization and  the Kerr switching occurs by switching the electric polarization P, or the antiferromagnetic order parameter L. Our findings suggest that 2D HS exhibiting ferroelectricity and antiferromagnetism could provide a novel material platform with  potential for information storage in antiferro-spintronics. 
 
\section{Acknowledgement}
We thank Ke Yang of University of Shanghai for Science and Technology, Xue-Zeng Lu, Haoshen Ye and Kaidi Xu of Southeast University for fruitful discussions. This work was supported by the National Natural Science Foundation of China (Grant Nos. 12274069 and 11834002), and Postgraduate Research \& Practice Innovation Program of Jiangsu Province (Grant No. KYCX22\underline{~}0241). We acknowledge the Project HPC-EUROPA3 (INFRAIA-2016-1-730897), with the support of the EC Research Innovation Action under the H2020 Programme, in particular, the computer resources and technical support provided by CINECA-HPC and the Big Data Center of Southeast University for providing the facility support on the calculations. Research at SPIN-CNR and University of L'Aquila has been funded by the European Union-NextGenerationEU under the Italian Ministry of University and Research (MUR) National Innovation Ecosystem grant ECS00000041-VITALITY. A. S. and C. R. acknowledges Università degli Studi di Perugia and MUR for support within the project Vitality.
	
\begin{suppinfo}

The SI is available free of charge on the ACS Publications website at DOI: 

More details about structural information of different stacking modes, electrical properties and Kerr angle of B-type CrI$_3$/In$_2$Se$_3$/CrI$_3$ HS.

\end{suppinfo}

\bibliography{achemso-demo}

\end{document}